\documentstyle[psfig]{mn}

\newif\ifAMStwofonts

\ifoldfss
  \ifCUPmtlplainloaded \else
    \NewTextAlphabet{textbfit} {cmbxti10} {}
    \NewTextAlphabet{textbfss} {cmssbx10} {}
    \NewMathAlphabet{mathbfit} {cmbxti10} {} 
    \NewMathAlphabet{mathbfss} {cmssbx10} {} 
  \fi
  \ifAMStwofonts
    \ifCUPmtlplainloaded \else
      \NewSymbolFont{upmath} {eurm10}
      \NewSymbolFont{AMSa} {msam10}
      \NewMathSymbol{\upi}     {0}{upmath}{19}
      \NewMathSymbol{\umu}     {0}{upmath}{16}
      \NewMathSymbol{\upartial}{0}{upmath}{40}
      \NewMathSymbol{\leqslant}{3}{AMSa}{36}
      \NewMathSymbol{\geqslant}{3}{AMSa}{3E}

    \fi
  \fi
\fi 

\ifnfssone
  \newmathalphabet{\mathit}
  \addtoversion{normal}{\mathit}{cmr}{m}{it}
  \addtoversion{bold}{\mathit}{cmr}{bx}{it}
  \newmathalphabet{\mathbfit} 
  \addtoversion{normal}{\mathbfit}{cmr}{bx}{it}
  \addtoversion{bold}{\mathbfit}{cmr}{bx}{it}
  \newmathalphabet{\mathbfss} 
  \addtoversion{normal}{\mathbfss}{cmss}{bx}{n}
  \addtoversion{bold}{\mathbfss}{cmss}{bx}{n}
  \ifAMStwofonts
    \ifCUPmtlplainloaded \else
      \UseAMStwoboldmath
      \makeatletter
      \new@mathgroup\upmath@group
      \define@mathgroup\mv@normal\upmath@group{eur}{m}{n}
      \define@mathgroup\mv@bold\upmath@group{eur}{b}{n}
      \edef\UPM{\hexnumber\upmath@group}
      \new@mathgroup\amsa@group
      \define@mathgroup\mv@normal\amsa@group{msa}{m}{n}
      \define@mathgroup\mv@bold\amsa@group{msa}{m}{n}
      \edef\AMSa{\hexnumber\amsa@group}
      \makeatother
      \mathchardef\upi="0\UPM19
      \mathchardef\umu="0\UPM16
      \mathchardef\upartial="0\UPM40
      \mathchardef\leqslant="3\AMSa36
      \mathchardef\geqslant="3\AMSa3E
    \fi
  \fi
\fi 

\ifnfsstwo
  \DeclareMathAlphabet{\mathbfit}{OT1}{cmr}{bx}{it}
  \SetMathAlphabet\mathbfit{bold}{OT1}{cmr}{bx}{it}
  \DeclareMathAlphabet{\mathbfss}{OT1}{cmss}{bx}{n}
  \SetMathAlphabet\mathbfss{bold}{OT1}{cmss}{bx}{n}
  \ifAMStwofonts
    \ifCUPmtlplainloaded \else
      \DeclareSymbolFont{UPM}{U}{eur}{m}{n}
      \SetSymbolFont{UPM}{bold}{U}{eur}{b}{n}
      \DeclareSymbolFont{AMSa}{U}{msa}{m}{n}
      \DeclareMathSymbol{\upi}{0}{UPM}{"19}
      \DeclareMathSymbol{\umu}{0}{UPM}{"16}
      \DeclareMathSymbol{\upartial}{0}{UPM}{"40}
      \DeclareMathSymbol{\leqslant}{3}{AMSa}{"36}
      \DeclareMathSymbol{\geqslant}{3}{AMSa}{"3E}
    \fi
  \fi
\fi 

\ifCUPmtlplainloaded \else
  \ifAMStwofonts \else 
    \def\upi{\pi}
    \def\umu{\mu}
    \def\upartial{\partial}
  \fi
\fi
\def\etal{{\it et al. }}

\begin{document}

\title[
Red Globular Clusters
]
{
The Connection between Globular Cluster Systems and the Host Galaxies
}
\author[
Duncan A. Forbes and Juan. C. Forte
]
{
Duncan A. Forbes$^{1,2}$ and Juan C. Forte$^{3}$\\
$^1$School of Physics and Astronomy, 
University of Birmingham, Edgbaston, Birmingham B15 2TT \\
(E-mail: forbes@star.sr.bham.ac.uk)\\
$^2$Astrophysics \& Supercomputing, Swinburne University,
Hawthorn VIC 3122, Australia\\
$^3$ Facultad de Ciencias Astronomicas y Geofisicas, Universidad
Nacional de La Plata, Paseo del Bosque, 1900 La Plata, 
Argentina\\ 
and CONICET
\\
(E-mail: forte@fcaglp.fcaglp.unlp.edu.ar)\\
}

\pagerange{\pageref{firstpage}--\pageref{lastpage}}
\def\LaTeX{L\kern-.36em\raise.3ex\hbox{a}\kern-.15em
    T\kern-.1667em\lower.7ex\hbox{E}\kern-.125emX}

\newtheorem{theorem}{Theorem}[section]

\label{firstpage}

\maketitle

\begin{abstract}

A large number of early type galaxies are now known to possess
blue and red 
subpopulations of globular clusters. We have compiled a database of 
28 such galaxies exhibiting bimodal globular cluster 
colour distributions. After converting to a common V--I colour
system, we investigate correlations between the mean colour of
the blue and red subpopulations with galaxy velocity dispersion.
We support previous claims 
that the mean colour of the blue globular
clusters are unrelated to their host galaxy. They must have
formed rather independently of the galaxy potential 
they now inhabit. The mean blue colour is similar to that for halo
globular clusters in our Galaxy and M31. 
The red globular clusters, on the other hand, 
reveal a strong correlation with galaxy 
velocity dispersion. Futhermore, in well--studied galaxies the 
red subpopulation has similar, and possibly
identical, colours to the galaxy halo stars. Our results indicate an 
intimate link between the red globular clusters and the host
galaxy; they share a common formation history. A natural
explanation for these trends would be 
the formation of the red globular clusters during galaxy collapse. 

\end{abstract}


\section{Background Motivation}

Globular clusters are 
relatively homogeneous entities containing stars of a
single age and metallicity. Although we do not fully understand the
relative efficiency of forming globular cluster (GC) stars over galactic
field stars, there is good evidence that when a starburst occurs in a
galaxy, GCs will also form (e.g. Ho 1997). 
Thus GCs can provide a unique probe of the star
formation and chemical enrichment history of galaxies. 
To fully exploit this `probe' one needs to have an idea for 
how GCs form in
galaxies. Current scenarios include mergers (Schweizer 1987; 
Ashman \& Zepf 1992), two phase {\it in
situ} (Forbes, Brodie \& Grillmair 
1997) and tidal stripping/accretion (Forte, Martinez \& Muzzio 1982; 
Cote, Marzke \& West 1998). 

A direct comparison of GC and host galaxy colours was first attempted in the late 
1970s and early 1980s (e.g. Hanes 1977; 
Forte, Strom \& Strom 1981). A difference in colour was
observed, in the sense that GCs were $\sim$0.5 dex more metal--poor than
galaxy stars. This lead Forte \etal (1981) to conclude that ``...the
chemical enrichment history of the globular cluster system and of the
spheroidal component must have 
differed, and that the globular clusters are
likely to form a component dynamically as well as chemically distinct from
the spheroid.''

However another line of argument has developed indicating 
that there {\it is} a 
link between GCs and their host galaxy. This involves the  
correlation between the mean metallicity of the GC system and galaxy
luminosity. Such a trend was first suggested by van den Bergh (1975) and
shown by Brodie \& Huchra (1991), but for only 10 galaxies. It was later
disputed by Ashman \& Bird (1993) claiming that there was no correlation,
and by Perelmuter (1995) claiming that the relation only existed for
spirals. In 1996, Forbes \etal showed a strong trend over 
10 magnitudes for
35 early type 
galaxies. Subsequent larger samples have confirmed a correlation (e.g. 
Durrell \etal 1996). Although the scatter is large, a linear fit has a
slope very similar to that for the galaxy stellar metallicity versus
galaxy luminosity relation, 
i.e. Z $\propto$ L$^{0.4}$ (see Brodie \& Huchra 1991). 

The galaxy relations, like the Mg$_2$ -- $\sigma$ and
the colour -- magnitude relations, 
are generally believed to be correlations of 
metallicity and mass 
(Kodama \& Arimoto 1997; 
Forbes, Ponman \& Brown 1998). Thus by analogy, the
mean 
metallicity of the GC system `knows about' the mass of the galaxy in which
it resides; {\it the chemical enrichment of GCs is directly linked to the 
galaxy formation process. }

It wasn't until 
the 1990s that we could easily
understand the 
apparent metallicity difference between the GC system and the 
host galaxy (as noted by Forte \etal 1981) -- it is due to the
presence of two
distinct GC subpopulations 
(e.g. Lee \& Geisler 1993).  The metallicity
difference being due to the relative mix of metal--rich and metal--poor
GC subpopulations. 
It appears that the metal--rich GCs have a mean metallicity that is similar
to the host galaxy stars with 
the metal--poor ones $\sim$1 
dex lower (e.g. Geisler \etal 1996; Forbes, Brodie \& Huchra
1997). In the Forte \etal (1981) study, their sampling was
dominated by the blue GCs leading them to suggest that the GC
system and the host galaxy were `distinct'. 

Revisiting the metallicity -- luminosity relation, 
Forbes \etal (1997) showed that for 11 GC
systems the metallicity of the metal--rich population correlated with
galaxy luminosity, but the metal--poor one did not (see also Burgarella, 
Kissler--Patig \& Buat 2000). Forbes \etal (1997) concluded that 
only the metal--rich GCs were closely linked to the 
formation process of the galaxy. 

However one 
caveat about the interpretation of the GC metallicity -- galaxy
luminosity relation is that metallicities are usually based 
on colours that
have been converted into [Fe/H] assuming a Galactic relation. This has two
limitations:\\
i) The Galactic relation is strictly only valid for very old GCs. The
limited current evidence suggests that GCs around ellipticals are indeed
very old (Kissler--Patig \etal 1998; Cohen, Blakeslee \& Rhyzov 
1998; Puzia \etal 1999), 
but this may not always be the case.\\
ii) Few Galactic GCs have solar or 
supersolar metallicities, whereas this is 
often the case for elliptical galaxy GCs. 
Recently a conversion from V--I to 
supersolar metallicities has been derived by 
Kissler--Patig \etal (1998), but it is still somewhat uncertain for
other colour systems. 

So although it seems fairly probable 
that the metal--rich GCs formed from gas that
has been processed within the potential well of the galaxy,
previously arguments relied on the assumption that the Galactic 
relation is applicable. 
In this paper, we use direct measurements of GC colours and
compare these to galaxy internal velocity dispersions for a
sample of 28 galaxies. 
Our approach alleviates some of the 
problems associated with metallicity -- luminosity 
relations mentioned above. In addition, we avoid the distance
dependence present in the metallicity -- luminosity relation. 
We also examine the GC and galaxy halo colours in detail for 
a few individual galaxies. 
We find that the mean colour of the red GCs correlates with
galaxy velocity dispersion. Furthermore, in well--studied
galaxies the mean colour of the red GCs is almost identical 
to the galaxy halo (bulge) stars.
We briefly discuss the implications of our findings for 
GC formation scenarios. 

\section{Conversion to a Single Colour System}

The bulk of high quality GC colours, 
available in the literature, come from {\it Hubble Space
Telescope} studies, and most of
these use the F555W (V) and F814W (I) filters. Although this choice of
filters does not provide as much metallicity `leverage' as say
B--I or C--T1
(on the Washington system), a large number of early type galaxies
are now known to possess GC systems with bimodal V--I 
colour distributions. Several more are known from observations in 
other colour systems. Before compiling a homogeneous dataset of
GC colours we need to 
convert the bimodality seen using 
other colour systems into V--I.  
We will use a V--I vs colour relation based on Galactic GCs, however 
we first need to confirm that it is applicable 
for the GC systems of early type galaxies.  

In Fig. 1 we show the extinction corrected (V--I)$_o$ versus 
(C--T1)$_o$ and (B--I)$_o$ 
colour for
Galactic GCs 
with low reddenings (i.e. E(B--V) $<$ 0.1) taken from Reed,
Hesser \& Shaw 
(1988) and Harris \& Canterna (1977). The best bisector fits 
are shown by a solid lines. We have also made a fit 
to transform B--R to V--I (not shown in Fig. 1). 
The fits are the following:\\

\noindent
(V--I)$_o$ = 0.49(C--T1)$_o$ + 0.32 \\
(V--I)$_o$ = 0.51(B--I)$_o$ + 0.11 \\
(V--I)$_o$ = 0.68(B--R)$_o$ + 0.15 \\

\noindent
The typical rms about the fits is $\pm$ 
0.03 mag. 
Thus one could expect to estimate V--I for a Galactic type GC 
with an accuracy of $\sim0.03$
mag. from a different colour.

\begin{figure*}
\centerline{\psfig{figure=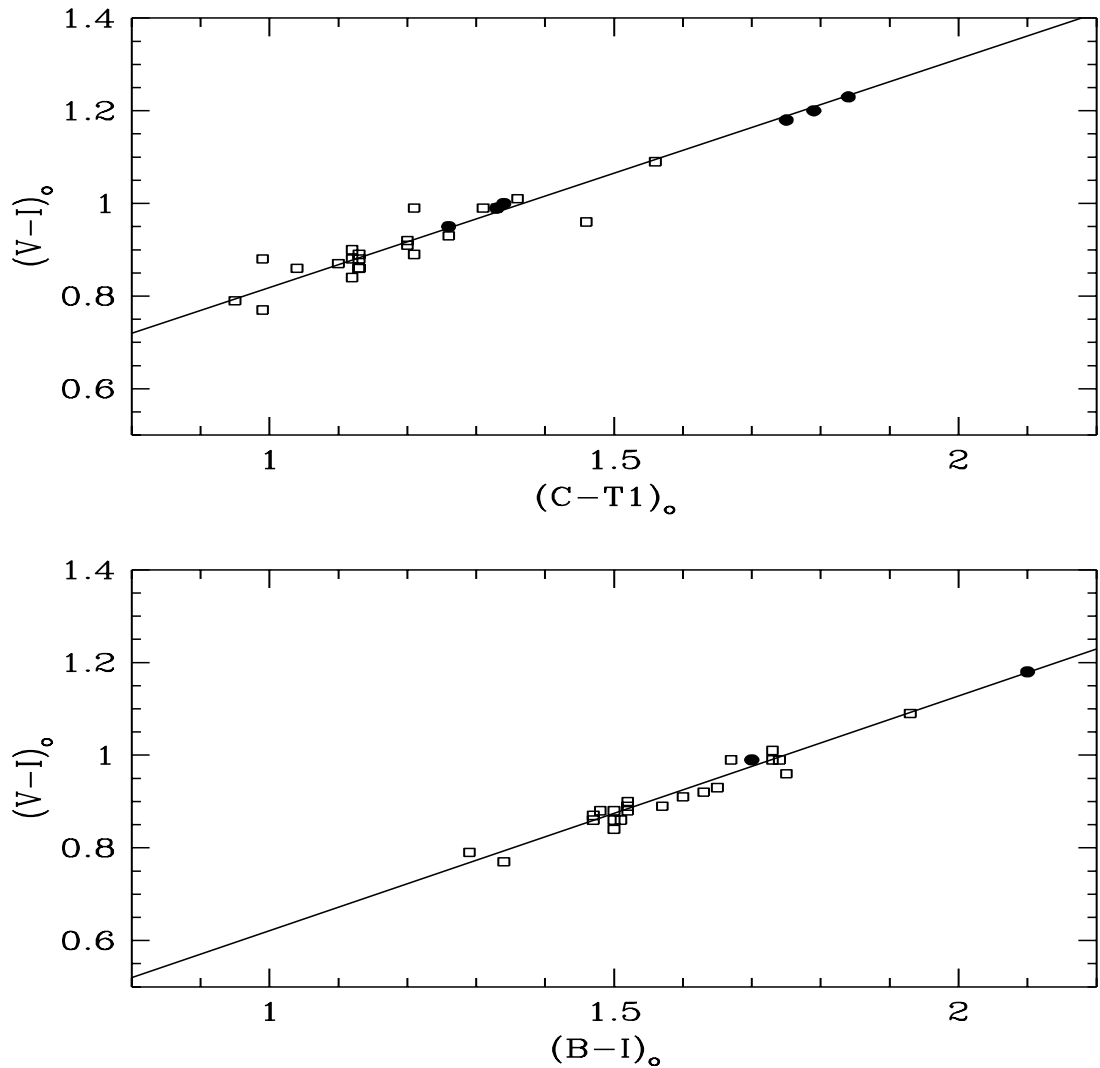,width=6in,height=6in}}
\caption{\label{fig1} 
(V--I)$_o$ vs (C--T)$_o$ and (B--I)$_o$ colour
relation for Galactic globular clusters. Open squares represent 
Galactic globular clusters with low reddening, and filled circles 
the blue and red globular cluster populations of NGC 1399,  
4472 and NGC 4486. The solid line is a fit to the Galactic
globular clusters only. 
}
\end{figure*}

Fig. 1 also shows the mean blue and red
colours for the GCs in the well--studied galaxies NGC 1399, 
NGC 4472 (M49) and NGC 4486 (M87). Here we have plotted the 
(C--T1)$_o$ measurements for NGC 1399 (Ostrov, Forte \& Geisler 1998), 
NGC 4472 (Geisler, Lee \& Kim 1996), and NGC 4486 (Lee \& Geisler 
1993) and the (B--I)$_o$ values for NGC 1399 (Forbes \etal 1998). The
corresponding (V--I)$_o$ values are NGC 1399 (Kissler--Patig \etal
1997), NGC 4472 (Puzia \etal 1999) and NGC 4486 (Kundu \etal
1999). 
The GC systems of these
galaxies are consistent with the Galactic
relation, including an extrapolation to redder (more metal--rich) 
colours.
Thus we feel confident in our transformation from C--T1 and B--I into
V--I. 
We do not have any independent confirmation of the B--R 
transformation but have no reason to doubt that it too is  
applicable to GCs in ellipticals. 

In Table 1 we list the mean 
(V--I)$_o$ colours for the GC subpopulations in 28 galaxies. Of
these, only four have been 
transformed from other colour systems (as noted in Table 1). 
The estimated
photometric errors for the mean colours are typically $\pm$
0.05 mag. (this includes the uncertainty of transformation
from one colour system to V--I). The confidence of bimodality is
given by `Yes' for definite and `Likely' for probable. In most
cases the statistical significance of the bimodality can be found 
in the original reference. 

\section{Results and Discussion}

\subsection{Trends with Velocity Dispersion}

In Fig. 2 we show the mean (V--I)$_o$ colour of the GC subpopulations 
versus the galaxy velocity dispersion for all of the galaxies
listed in Table 1. The velocity dispersion data come from 
Prugniel \& Simien (1996). 

The blue GCs show no correlation with galaxy velocity dispersion. 
Indeed they reveal a fairly constant colour of 
(V--I)$_o$ = 0.954 $\pm$ 0.008. This is very
similar to the overall mean colours of the Milky Way (0.94 $\pm$
0.01) and M31 (0.96 $\pm$ 0.01) GC systems (see Barmby \etal
2000). If we exclude the disk/bulge populations of GCs in 
these spirals, then the
halo GCs in both galaxies have a 
mean (V--I)$_o$ $\sim$ 0.92 (Barmby \etal 2000). 
Thus the blue GCs in 
spirals and early type galaxies have a very 
similar age and metallicity, but there are hints that those in
spirals may be slightly younger and/or more metal--poor. 

\begin{figure*}
\centerline{\psfig{figure=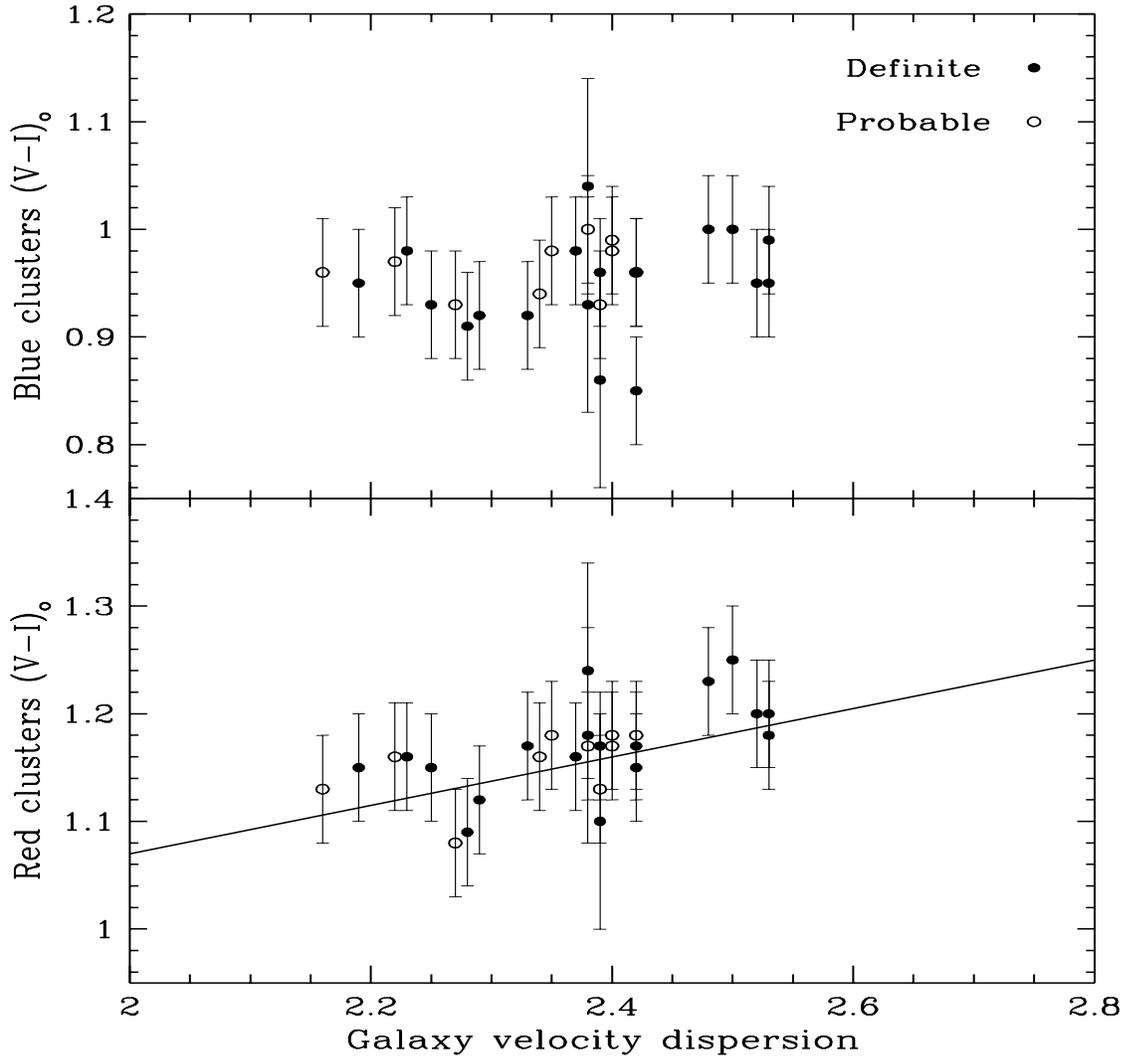,width=6in,height=6in}}
\caption{\label{fig2} 
Colour of globular cluster subpopulations versus log of the 
galaxy velocity
dispersion. 
Solid circles are galaxies with
definite bimodal globular cluster colours, and open circles are
probable bimodal systems. A 3$\sigma$ correlation exists between globular
cluster mean colour and velocity dispersion for the red
subpopulation but not for the blue subpopulation. 
}
\end{figure*}

The red GCs, on the other hand, reveal a strong correlation 
(at 3$\sigma$ significance) of mean colour with 
galaxy velocity dispersion.
The best fit 
line for the red GCs (V--I = 0.23~log~$\sigma$ + 0.61)  
is shown in Fig. 2. 
Galaxy velocity dispersion is a
tracer of galaxy mass. Thus the colour (and presumably
metallicity) of the red GCs is directly linked to the depth of
the galaxy's potential well. 

We note that if we had converted the C--T1 measurements of Secker 
\etal (1995) for NGC 3311 
then the colours of the blue (V--I = 1.15) and red (V--I = 1.28)
GCs would deviate strongly from the trends seen in
Fig. 2. However we used the more recent results of Brodie, Larsen 
\& Kissler--Patig 
(2000) which indicate normal GC colours for NGC 3311. We suspect
zero point errors in the Secker \etal (1995) analysis. This may
also affect the C--T1 colours (Zepf \etal
1995) for the NGC 3923 GC system which was observed on the same run
(we have assigned it a larger error). 
Indeed NGC 3923 is one of the more 
deviant points in Fig. 2. 

\subsection{Trends with Galaxy Halo Colour}

Having shown that the red GC subpopulation has a direct
connection with the host galaxy, while the blue subpopulation is
unconnected we now explore the trend between mean GC colour and
galaxy halo colour. 

In order to best compare GC and galaxy colours, the 
colours should come from the same observations, 
and sample a similar
galactocentric annulus around any given galaxy as radial
gradients may exist in the galaxy and GC subpopulations (e.g. 
Ostrov \etal 1998; Forte \etal 2000). We have also decided to
restrict our analysis to the wide--area studies conducted using 
C--T1 (which has the best metallicity sensitivity of GC
studies). This excludes HST studies which typically probe only the
galaxy inner regions in V--I. 

With these criteria we found four galaxies 
(the Secker \etal (1995) C--T1 data on NGC 3311 has been excluded
for reasons mentioned above).  
The four are NGC 1399 (Ostrov \etal 1998), NGC 1427
(Forte \etal 2000), 
NGC 3923 (Zepf \etal 1995), NGC 4472
(Geisler \etal 1996; Lee \& Kim 2000) and NGC 4486 (Geisler \etal
2000). We note that the zero points and hence colours of 
NGC 3923 may be in error (as discussed above) but
as the galaxy and GCs colours come from the same observation this 
will not effect the relative colour difference.
The two papers referenced for NGC 4472 use the same data
collected in 1993. 
Where possible, colours have been taken from similar 
galactocentric radii. In the case of NGC 1399, Ostrov \etal
(1998) quote galaxy and GC colours in three radial bins for the
red subpopulation and two bins for the blue ones. 

\begin{figure*}
\centerline{\psfig{figure=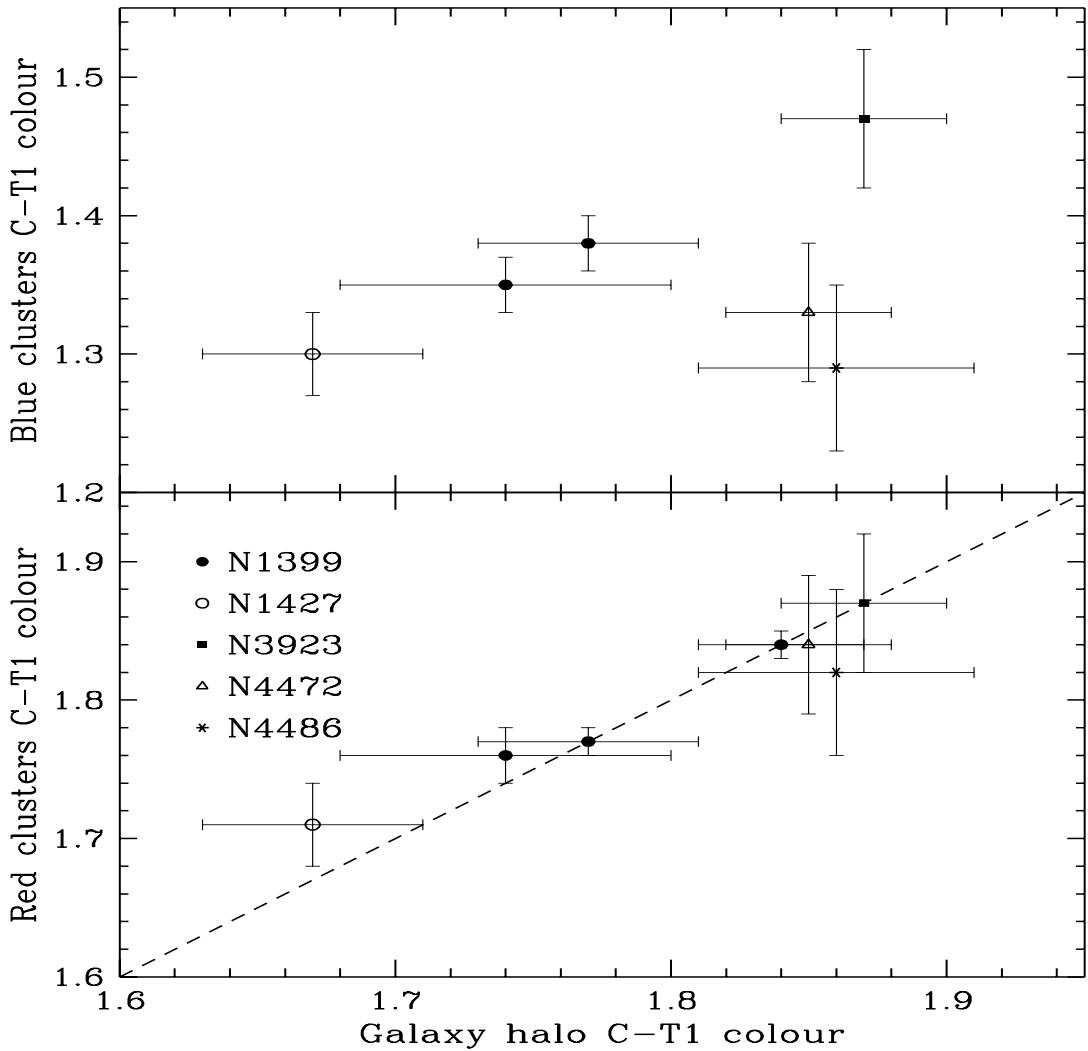,width=6in,height=6in}}
\caption{\label{fig3} 
Mean C--T1 colour of the globular clusters and galaxy halo.
Globular cluster and galaxy colours are taken from the same
source, and matched in galactocentric radius where possible.
For NGC 1399 colours at different galactocentric radii are shown. 
See text for details. 
The dashed line shows a one-to-one relation. 
Within the errors,  
red globular clusters have identical C--T1 colours to the host 
galaxy halo.  
}
\end{figure*}

The mean GC colours versus galaxy halo colour is shown in
Fig. 3. The mean blue GC colours do not reveal a strong trend
with halo colour, and in particular lie well away from a
one-to-one relation. However the red GCs {\it do} lie close to a
one-to-one relation. 
Within the errors, the red GCs have identical C--T1 colours
to those of their host galaxy. So not only do red GCs have a
metallicity that is linked to the depth of the galaxy's
potential well (Fig. 2), for at least {\it some} galaxies 
they also have the same metallicity
as the galaxy halo stars. This suggests a very well coordinated
star formation and chemical enrichment history. 

      For the NGC 1399 red GCs, the agreement is not only
      restricted to the mean colours discussed above, but also in
      the sense that they share the same colour gradient as the
      underlying galaxy halo (see table 5 in Ostrov \etal 1998).
      If the galaxy halo was in fact 
related to the {\it overall} GC system, then halo
colour gradients would reflect the varying ratio of the blue to
red GCs with galactocentric radius. However if this were the
case, we would expect the halo colour gradient to get
significantly bluer with
increasing radius (as the blue GCs dominate).  
For NGC 1399 (Ostrov \etal 1998) this is not the case indicating that the 
halo colours do not reflect the blue GCs, but rather the
red ones. 

Finally, we note that in recent ground--breaking work, 
Harris \etal (1999) used HST to resolve the stars in the halo of
NGC 5128. They showed that the red halo 
stars and the metal--rich GCs have the same metallicity, and
concluded that they formed in 
the same star formation event. 

\subsection{How did the Red Globular Clusters Form ?}

The simplest explanation for the fact that the red GCs and
the galaxy stars may have similar colours, and
that red GCs colours correlate with galaxy velocity dispersion, is that
they formed at the same time from the same chemically enriched
gas. This would naturally favour the two phase 
collapse scenario (Forbes \etal
1997). In this scenario the red GCs form almost contemporaneously 
with the galaxy field stars, and hence share their
properties. The trends described here are a natural consequence
of this idea. 

Are these trends expected in the merger picture (Ashman \& Zepf
1992) ? After a gaseous merger the galaxy will consist of 
at least two stellar and two GC populations 
(the first associated with the progenitor galaxies
and the other created in the merger). The relative ratios of 
these two components depend on how much gas is available to
create the new stars and GCs. The gas fraction would be high at 
early epochs. 
To explain Fig. 2 one could 
argue that the more metal--rich (redder) GCs formed in the larger
(higher velocity dispersion) galaxies. However to explain the
nearly identical colours of the galaxy halo and red GCs (Fig. 3), 
one must 
then argue that the halo is dominated by new stars formed at the
same time, from the same gas, as the new (red) GCs. This would
imply that the progenitors were largely gaseous and hence the
merger occurred at high redshift. The distinction between the
merger and collapse pictures becomes blurred. 
Various other difficulties for the merger 
model are discussed in detail by Forbes \etal (1997). 

In the accretion model of Cote \etal (1998), the red GCs form
{\it in situ} but the blue ones are accreted, along with
starlight, from stripped dwarf galaxies. This accumulation of
metal--poor material into the halo of the primary galaxy would
tend to make the halo light bluer than the red GCs -- in
contradiction to what is seen in Fig. 3. 

\section{Concluding Remarks}

Here we have presented evidence that the mean colour of the 
blue GCs in early type galaxies are
unrelated to their host galaxy, supporting the work of Forbes
\etal 
(1997) and Burgarella \etal (2000). 
In terms of a collapse scenario, where 
the blue GCs form in a proto--galactic dark matter halos, they do not
appear to have `memory' of the collapse phase suggesting they formed
pre--collapse or possibly quite independently of the eventual
host galaxy (Burgarella \etal
2000). We note however, that this is probably not true for 
the Milky Way halo globular clusters (the `blue' subpopulation). 
These metal--poor globular clusters 
{\it do} show evidence that the majority formed within our
Galactic halo (van den Bergh 1996). 

The red GCs, on the other hand, do `know' about the galaxy
they occupy in terms of its potential well and halo colour. 
This suggests that the red GCs and the host galaxy 
share a common formation event. This naturally 
favours the two phase collapse
picture, in which the red GCs and the halo stars are formed
together (Forbes \etal 1997). The merger picture (Ashman \& Zepf
1992) is only compatiable if the merger took place at early
epochs involving gaseous progenitors. \\

\noindent{\bf Acknowledgements}\\
We thank A. Georgakakis, S. Larsen and S. van den Bergh  
for useful discussions. S. Larsen also kindly provided results
on NGC 3311 before publication. DAF thanks La Plata Observatory for their
hospitality during August 1999 when much of work was carried
out. Finally we thank the referee J. Brodie for her many 
useful comments and suggestions.\\

\noindent{\bf References}

\noindent
Ashman, K.M., Zepf S.E. 1992, ApJ 384, 50 \\
Ashman K.M., Bird C.M., 1993, AJ, 106, 2281\\
Barmby, P., Huchra, J.P., Brodie, J.P., Forbes, D.A., Schroder,
L.L., Grillmair, C.J., 2000, AJ, 119, 727\\
Brodie, J.P., Huchra, J.P., 1991, ApJ, 379, 157\\
Brodie, J.P., Larson, S., Kisser--Patig, M., 2000, ApJ, submitted\\
Burgarella, D., Kissler--Patig, M., Buat, V., 2000, A\&A, in press\\
Cohen J.G., Blakeslee J.P., Rhyzov A. 1998, ApJ 496, 808\\
Cote, P., Marzke, R.O., West, M.J., 1998, ApJ, 501, 554\\
Forbes D.A., Franx M., Illingworth G.D., Carollo C.M. 1996, ApJ 467, 126\\
Forbes D.A., Brodie J.P., Grillmair, C.J. 1997, AJ 113, 1652 \\
Forbes D.A., Grillmair, C.J., Williger, G.M., Elson, R.A.W.,
Brodie, J.P., 1998, MNRAS, 293, 325\\
Forbes, D.A., Ponman, T.J., Brown, R.J.N., 1998, 508, L43\\
Forbes D.A., Brodie J.P., Huchra J. 1997, AJ 113, 887 \\
Forte, J.C., Strom, S.E., Strom, K.M., 1981, ApJ, 245, L9\\
Forte, J.C., Martinez, R.E., Muzzio, J.C., 1982, AJ, 87, 1465\\
Forte, J.C., Geisler, D., Ostrov, P.G., Piatti, A., 2000, in
prep.\\ 
Gebhardt, K., Kissler--Patig, M., 1999, AJ, 118, 1526\\
Geisler D., Lee M.G., Kim E., 1996, AJ 111, 1529 \\
Geisler D., \etal 2000, in prep.\\
Hanes, D., 1977, MNRAS, 179, 331\\
Harris, H., Canterna, R., 1977, AJ, 82, 798\\
Harris, G., Harris, W.E., Poole, G.B., 1999, AJ, 117, 855\\
Ho, L., 1997, Rev. Mex. Astr. Astro., 6, 5\\ 
Kim, E., Lee, M. G., Geisler, D., 2000, MNRAS, in press\\
Kissler-Patig M., Kohle S., Hilker M., Richtler T., Infante, L., Quintana
H., 1997a, A\&A 319, 470\\
Kissler-Patig M., Richtler T., Storm M., Della Valle M., 1997b, A\&A 327,
503\\
Kissler-Patig M., Brodie J.P., Forbes D.A., Grillmair C.J., Huchra J.A.
1998, AJ 115, 105\\
Kodama, T., Arimoto, N., 1997, A\&A, 320, 41\\
Kundu, A., Whitmore, B.C., 1998, AJ, 116, 2841\\
Kundu, A., Whitmore, B.C., Sparks, W.B., Machetto, F.D., Zepf,
                   S.E., Ashman, K.M., 1999, AJ, 515, 733\\
Kundu, A., 1999, PhD Thesis, University of Maryland\\
Larson, S., Brodie, J.P., 2000, AJ, submitted\\
Lee M.G., Geisler D., 1993, AJ 106, 493 \\
Ostrov, P.G., Forte, J.C., Geisler, D., 1998, AJ, 116, 2854\\
Prugniel, P., Simien, F., 1996, A\&A, 309, 749\\
Perelmuter, J.M., 1995, ApJ, 454, 762\\
Puzia, T., Kissler--Patig, M., Brodie, J.P., Huchra, J.P., 1999,
AJ, 118, 2734\\
Reed, B. C., Hesser, J.E., Shawl, S.J., 1988, PASP 100, 545\\
Secker, J., Geisler, D., McLaughlin, D.E., Harris, W.E., 1995,
AJ, 109, 1019\\
Schweizer, F., 1987, in Nearly Normal Galaxies, ed S. Faber (Springer, New
York), 18\\
Van den Bergh S., 1975, ARAA, 13, 217\\
Van den Bergh S., 1996, PASP, 108, 986\\
Woodworth, S.C., Harris, W.E., 2000, astro-ph/0002292\\
Worthey G., 1994, ApJS 95, 107 \\
Zepf, S.E., Ashman, K.E., Geisler, D., 1995, ApJ, 443, 570

\end{document}